\documentclass{aa}
\usepackage{graphics}
\begin{document}

   \thesaurus{13(13.07.1;04.19.1)}
   \title{Gamma-Ray Bursts Detected From BATSE DISCLA Data}

   \author{M. Schmidt}

   \offprints{M. Schmidt}

   \institute{California Institute of Technology, Pasadena, CA 91125, USA\\
              email: mxs@deimos.caltech.edu}

   \date{Received January 21; accepted February 28, 1999}

   \maketitle

   \begin{abstract}

   We have searched for gamma-ray bursts (GRB) in the BATSE DISCLA data
   covering the period TJD 8365-10528. We employ an algorithm that uses
   the background both before and after the onset of the burst, and that
   requires an excess of at least $5 \sigma$ over background in the energy
   range $50-300$ keV. For the initial set of 7536 triggers, we find strong 
   concentrations at given geographic locations of the satellite. 
   After excluding these
   geographic areas, we are left with 4485 triggers. We exclude triggers
   close in position to solar flares, CygX-1, and Nova Persei 1992, when active
   as well as soft bursts close to the sun. We accept 1018 triggers that
   occur within 230 sec of a GRB listed in the BATSE catalog, 
   and upon visual inspection
   classify 881 triggers as magnetospheric events and 404 as GRBs. 
   The final sample of 1422 GRBs represents effectively 2.0 years of
   isotropic exposure, for an annual rate of 710 GRBs per year.


   \end{abstract}

%
%________________________________________________________________

\section{Introduction}

   The Burst and Transient Source Experiment (BATSE) on board the
   {\it Compton Gamma Ray Observatory} has been very effective in
   recording gamma-ray bursts (GRB). Soon after the observatory's launch
   on April 19, 1991, the results obtained with BATSE showed convincingly 
   that the sky distribution of GRBs was isotropic 
   (Meegan et al. 1992a) and that the line-of-sight distribution
   was incompatible with a homogeneous distribution in
   euclidean space (Meegan et al. 1992b). 
   The proposition based on these findings that the distance 
   scale of GRBs is cosmological (Paczynski 1992) was
   eventually confirmed by the observation of a large redshift 
   (Metzger 1997). 

   The GRB data from BATSE have been published primarily in a succession
   of catalogs (cf. Meegan et al. 1997). These data are
   based on an on-board trigger, which acts when certain conditions
   are fulfilled. Usually these require that the counts in the energy 
   range $50-300$ keV exceed the background by $5.5 \sigma$
   on a time scale of 64, 256 or 1024 msec in at least two of the
   eight detectors. Variations of these criteria, in the channels and
   the signal-to-noise limits used, have been in effect at various
   times (cf. Meegan et al. 1997).

   BATSE produces data in various modes in archival form. The DISCLA
   data provide a continuous record of the counts in channels 1, 2, 3,
   and 4 for the eight detectors on a time scale of 1024 msec. The
   DISCLA data allow the detection of GRBs {\it a posteriori} in a
   way similar to the on-board trigger (except for the availability of
   only one time scale).

   There are distinct advantages to searching for GRBs in archival data: 
   the detection parameters can be set independently of those that are
   hard-wired in the on-board trigger mechanism; the search can be
   repeated with a different detection algorithm; each search can be 
   carried out on the full data set, etc. A search for GRBs not detected
   by the on-board BATSE trigger has been conducted by Kommers et al. (1997).

   We have used the DISCLA data to search for GRBs in the time period TJD 
   $8365-10528$. In the following sections we describe the search,
   the classification of the triggers, and the derivation of $<V/V_{max}>$
   of the resulting sample.

\section{The search}
 
%----------------------------------------------------------- 
   \begin{figure}
   \resizebox{\hsize}{!}{\includegraphics{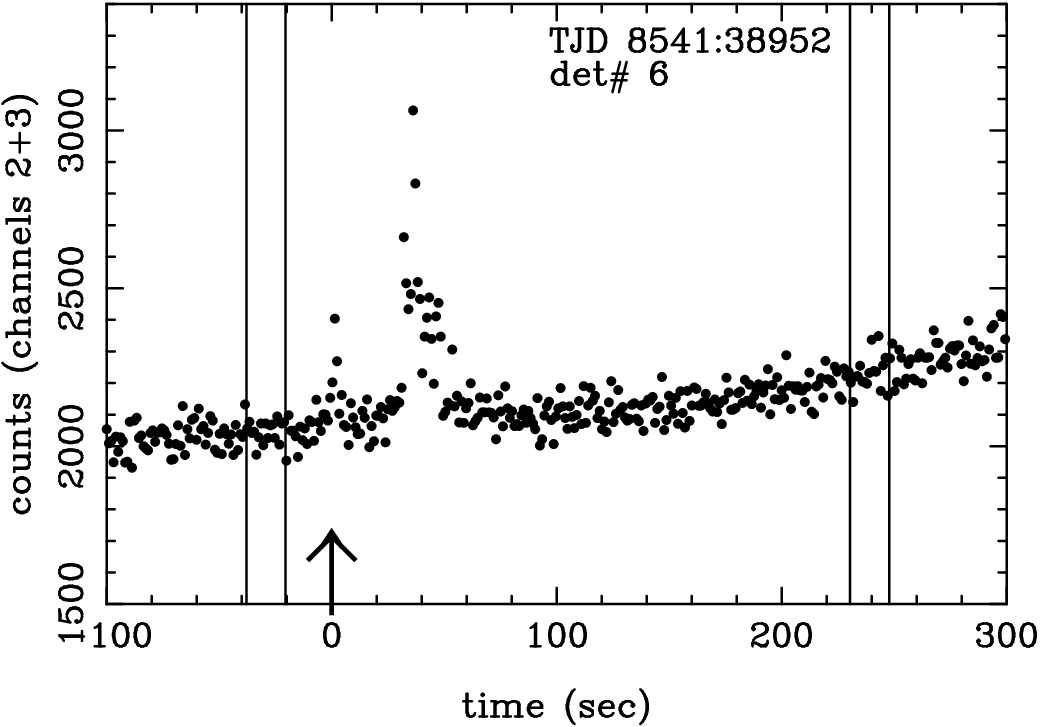}}
   \caption[]{The test for the presence of a burst at time $0.0$ 
    involves two intervals of 17.408 sec in which the background
    is derived, one ending at time -20.48 sec, the other beginning
    at time 230.4 sec, see text.}
   \label{bkgrd}
   \end{figure}
%
%______________________________________________________________
%__________________________________________________________________

   We have tried to automate the search procedure as much as possible,
   for the sake of both speed and consistency. This will allow simulations 
   to be carried out to investigate various scientific problems. 
   Our search has some similarities to that carried out by the
   on-board BATSE trigger: using the counts in channels $2+3$ covering
   the energy range $50-300$ keV; averaging background counts over
   17.408 s; and requiring that two modules see a minimum excess over
   background. We differ, however, in the evaluation of the background
   as explained below.

   \begin{figure}
   \resizebox{\hsize}{!}{\includegraphics{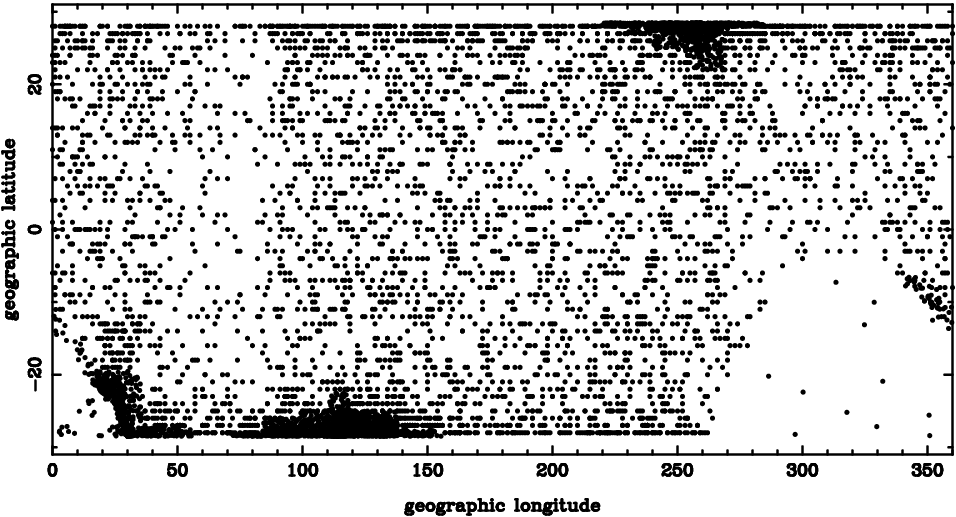}}
   \caption[]{Geographic positions of the Compton satellite during
    7536 triggers. The geographic areas in the denser parts of the
    plot were excluded from the search, eliminating 3051 triggers.}
   \label{geo}
   \end{figure}
%
%______________________________________________________________

   In the search conducted by the 
   on-board BATSE trigger on the 1024 msec time scale, 
   the background is derived over a given stretch of 17.408 sec and the
   trigger test is carried out for the 1024 msec bin following that stretch.
   The same background stretch is used for the next sixteen 1024 msec bins, so
   that there is a separation between the end of the background stretch and
   the test bin of $0 - 16.384$ sec.

   We use a fixed separation between the background stretch and the
   test bin, and also introduce a second background stretch allowing a
   linear interpolation of the background for the duration of the GRB,
   cf. Fig. 1. For a small or zero separation between background stretch 
   and test bin, slowly rising GRBs may escape detection (Higdon and 
   Lingenfelter 1996). If we increase the separation, more slowly rising
   GRBs can be detected, but then we find that the number of false triggers 
   caused by higher-order variations in the background increases 
   substantially. In the
   current search, we used a separation of 20.48 sec, and placed
   the beginning of the second background stretch 230.4 sec after the
   test bin, as shown in Fig. 1. Following a burst, we disabled the
   trigger mechanism for 230.4 sec.

%----------------------------------------------------------- 
   \begin{figure}
   \resizebox{\hsize}{!}{\includegraphics{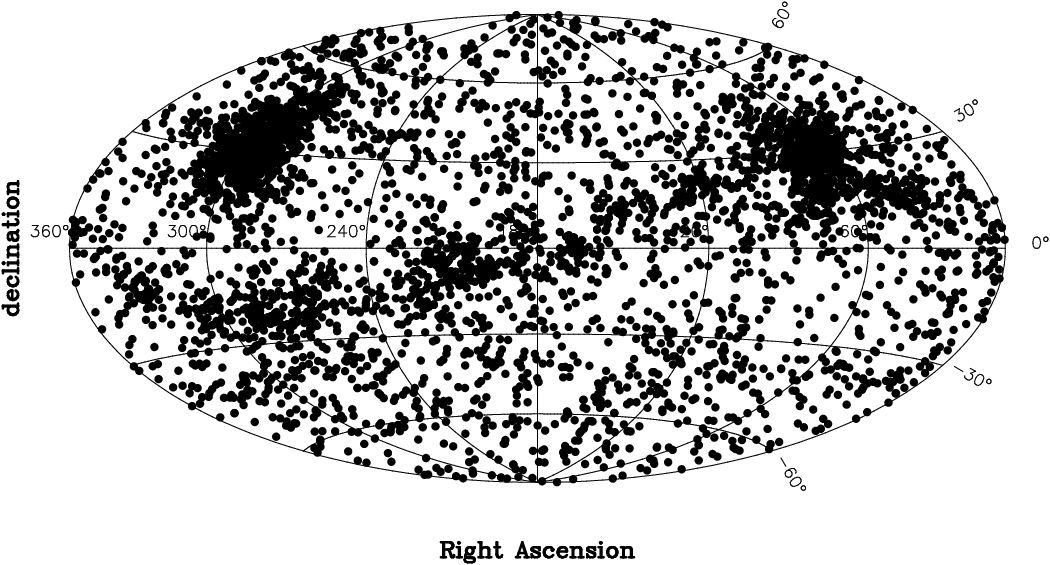}}
   \caption[]{Equatorial coordinates of 4485 triggers. The effects of 
   CygX-1, Nova Persei 1992, and solar flares along the ecliptic
   are clearly seen. The remaining background triggers are
   partly magnetospheric events, and partly GRBs.}
   \label{sky}
   \end{figure}
%
%______________________________________________________________

   Artifacts or defects in the data will lead to false triggers. We have
   systematically searched for gaps in the data, and for constant output 
   numbers (usually zeroes), and, from quality data provided by the BATSE 
   project, for the occurrence of checksum errors. 
   In each case, we set up a time window
   of exclusion around the defect, so that the automatic application of
   our search algorithm will not lead to a trigger. Considerable time
   is lost: the total number of our exclusion windows is over $151,000$,
   of which some overlap.

   We adopted a limiting $S/N$ ratio for detection in two detectors of 
   $5.0$. Our search of the time period TJD $8365-10528$ with the criteria
   described above yielded 7536 triggers. The geographic coordinates
   of the observatory at the time of trigger are plotted in Fig. 2.
   Besides some areas near the South Atlantic Anomaly, there are clear
   concentrations in the south over W. Australia and in the north over
   Mexico and Texas. The southern concentration has been discussed by
   Horack et al. (1992). Since these concentrations have nothing to do
   with cosmic GRBs, we outlined geographic exclusion regions to
   avoid most of these triggers. With these exclusions in place, we
   are left with 4485 triggers, which form the basis for classification
   and discussion.

\section{Classification of triggers}

   For each of the 4485 triggers, we derived sky positions on the
   assumption that they corresponded to the emission from a point source.
   We used the response function for the eight detectors (cf. Pendleton 
   et al. 1995), and assumed a Band et al. (1993) 
   type spectrum with $\alpha = -1$, $\beta = -2$, and
   break energy $E_0 = 150$ keV. The positions were derived by testing
   around $10,000$ positions on the sky and minimizing $\chi^2$ of the counts
   for all eight detectors. We ignored the effects of Compton scattering
   on the Earth's atmosphere or the spacecraft. Fig. 3 shows the celestial
   coordinates of all 4485 triggers.

   It is clear from an inspection of Fig. 3 that at least four types
   of triggers are present. Both CygX-1 and Nova Persei 1992 are
   prominent. The sun is clearly shown by solar flares along the ecliptic.
   There is a fourth component which appears more or less isotropic.

%__________________________________________________ One column table
   \begin{table}
      \caption[ ]{Classification of 4485 triggers}
      \label{class}
%      \[
%         \begin{array}{p{0.5\linewidth}l}
         \begin{flushleft}
         \begin{tabular}{lrr}
            \noalign{\smallskip}
            \hline
            \noalign{\smallskip}
            Description    &  reject  &  accept  \\
            \noalign{\smallskip}
            \hline
            \noalign{\smallskip}
            Within 23 deg of sun, when active   & 963 &   \\
            Within 23 deg of CygX-1, when active        & 827 &   \\
            Within 23 deg of Nova Per, when active & 418 &   \\
            Within 230 sec of BATSE-listed GRB  &   & 1018  \\
            Profile inspected: accepted as GRB      &   &  389  \\
            Profile inspected: probably a GRB     &   &   15  \\
            Profile inspected: rejected      & 881  &    \\
            Spectrum very soft, near sun: rejected &  44  &    \\
            \noalign{\smallskip}
            \hline
            \noalign{\smallskip}
            Total number of GRBs in DISCLA sample:  &   & 1422  \\
            \noalign{\smallskip}
         \end{tabular}
         \end{flushleft}  
%         \end{array}
%      \]
   \end{table}

%----------------------------------------------------------- 

   Plots of the number of triggers versus time close to the sky positions of
   CygX-1 and of Nova Persei 1992 show that both were detected during
   well-defined periods of activity. We used our
   positions of Nova Persei 1992, which totally dominated activity in
   its part of the sky for around 60 days, to evaluate the distribution
   of position errors. Based on these results, we eliminated from
   consideration all triggers within 23 deg of the sun, of CygX-1,
   and of Nova Persei 1992 during the periods that these sources were active, 
   see Table 1. 

   We accept as GRBs all triggers whose onset was within 230.4 
   sec of those listed for GRBs in the BATSE catalog (cf. Meegan et al. 1997
   and the World Wide Web version maintained by the BATSE Burst team).
   For the remaining triggers, we inspected each
   of their time profiles from $-300$ to $+400$ sec relative to the trigger
   time. In our judgment of the nature of these events we were guided by
   descriptions by the BATSE team of magnetospheric events (cf. Fishman
   et al. 1992, Horack et al. 1992),
   by the value of $\chi^2$ of the solution, by the dispersion of
   the positions obtained during each second that the burst was brighter
   than the limiting flux, and by a derivation of the angular motion of 
   the source (which for a GRB should be consistent with zero). 
   In the process, we noticed that 44 of the remaining triggers
   had spectra softer than any listed in the BATSE catalog, and that they
   were all within 35 deg of the sun. On the basis of this evidence, we
   rejected these as GRBs. The results of this excercise are shown in Table 1. 
   We accepted 404 GRBs that are not listed in the BATSE catalog.

   There are 130 GRBs listed in the BATSE catalog that we have not detected,
   while as stated we find 404 GRBs that are not in the catalog.
   The difference in content between the BATSE catalog and our sample
   is partly a consequence of differences in the de-activation of the trigger
   following a burst or around a defect in the data, and partly caused
   by statistical differences in the different backgrounds used.

   Based on the entire search procedure, including the time windows of
   exclusion set up around bad data, the exclusion of particular
   geographic areas and of some sky areas in the classification
   procedure, and the part of the sky occulted by the Earth, we estimate 
   that the sample of 1422 GRBs represents effectively 2.0 years of isotropic
   exposure by BATSE, for an annual detection rate of 710 GRBs per year.

\section{Derivation of $<V/V_{max}>$} 

   The most important property of the GRBs in the sample is the euclidean
   value of $<V/V_{max}>$. We will use its value to derive a distance
   scale for GRBs in a subsequent communication.

   In most evaluations of $V/V_{max}$ for individual GRBs, it has been 
   assumed that it equals $(C_{max}/C_{min})^{-3/2}$ where $C_{min}$ is the
   minimum detectable count rate and $C_{max}$ is the count rate at
   the maximum amplitude of the burst. This may not be correct. If we 
   remove the source to the largest distance at which it is just still 
   detectable, it is likely that the detection of the burst will occur
   later, and therefore that $C_{min}$ will include some burst signal.
   Also, in some cases the part of the burst containing $C_{max}$ will
   not cause a trigger when we remove the burst. Due to
   these inescapable effects, $V/V_{max}$ will be larger than follows
   from the values of $C_{max}$ and $C_{min}$ at detection. This
   situation was encountered by Higdon and Schmidt (1990) in their
   discussion of GRBs from the Venera 11 and 12 KONUS experiments.

   We have derived the euclidean value of $V/V_{max}$ for each GRB
   as follows. Upon detection, we derive the burst time profile by
   subtracting the background (interpolated between the two background
   stretches, see Fig. 1) from the counts. We multiply this original
   time profile by a factor of X and add it to the background. Next we
   apply the search algorithm to see whether we detect a burst. If so,
   we repeat the process with a smaller value of X, until we do not
   detect a burst anymore. This represents the situation (as best we can)
   when the source has been removed 
   in distance by a factor $X^{-1/2}$ in euclidean
   space. Therefore $V/V_{max}$ for the source is $X^{3/2}$. Application
   of this process to all bursts in the sample produces 
   $<V/V_{max}> = 0.334 \pm 0.008$.

\begin{acknowledgements}
During the many years in which this work came to fruition, I have had much
appreciated help from or discussions with D. Chakrabarty, M. Finger,
G. Fishman, J. Gunn, J. Higdon, J. Horack, C. Meegan, T. Prince, and
B. Vaughan. 

\end{acknowledgements}


\begin{thebibliography}{}

   \bibitem[1993]{band} Band D.L. et al., 1993
    ApJ 413, 281

   \bibitem[1992]{fishman} Fishman G.J., Meegan C.A., Wilson R.B., 
    et al., 1992,
      in: AIP Conf. Proc. 265, Gamma-Ray Bursts,
      eds.\ W. S. Paciesas, G. J. Fishman,
      AIP, New York, p.\ 13

   \bibitem[1996]{higdon} Higdon J.C., Lingenfelter R.E., 1996,
      in: AIP Conf. Proc. 384, Gamma-Ray Bursts,
      eds.\ C. Kouveliotou, M. F. Briggs, G. J. Fishman,
      AIP, New York, p.\ 402

   \bibitem[1990]{hsc} Higdon J.C., Schmidt M., 1990,
    ApJ 355, 13

   \bibitem[1997]{kommers} Kommers J.M., Lewin W.H.G., Kouveliotou C. 
    et al., 1997,
    ApJ 491, 704 

   \bibitem[1992]{iso} Meegan C.A., Fishman G.J., Wilson R.B., et al., 1992a,
    Nature 355, 143

   \bibitem[1992]{nonh} Meegan C.A., Fishman G.J., Wilson R.B., et al., 1992b,
      in: AIP Conf. Proc. 265, Gamma-Ray Bursts,
      eds.\ W. S. Paciesas, G. J. Fishman,
      AIP, New York, p.\ 61

   \bibitem[1997]{4b} Meegan C.A., Paciesas W.S., Pendleton G.N. et al., 1997,
      in: AIP Conf. Proc. 428, Gamma-Ray Bursts,
      eds.\ C. A. Meegan, R. D. Preece, T. M. Koshut,
      AIP, New York, p.\ 3
   
   \bibitem[1997]{metzger} Metzger M., et al., 1997
    Nature 387, 878

   \bibitem[1992]{paczynski} Paczynski B., 1992,
      in: AIP Conf. Proc. 265, Gamma-Ray Bursts,
      eds.\ W. S. Paciesas, G. J. Fishman,
      AIP, New York, p.\ 144

   \bibitem[1995]{pendleton} Pendleton G.N., Paciesas W.S.,
    Mallozzi R.S., et al., 1995,
    Nucl. Instrum. Methods Phys. Res. A 364, 567



\end{thebibliography}
\end{document}